\documentstyle{article}
\begin{document}
\newtheorem{Prop}{Proposition}
\newtheorem{Theor}{Theorem}
\newtheorem{Defi}{Definition}
\newtheorem{Lem}{Lemma}
\newtheorem{Res}{Result}

\font\Sets=msbm10
\def\Complex {\hbox{\Sets C}}
\def\Real {\hbox{\Sets R}}
\def\Nat {\hbox{\Sets N}}
\def\Integ {\hbox{\Sets Z}}

\title{Integrable ODEs on Associative Algebras}

\author{A.V.~Mikhailov\\
Applied Math. Department, University of Leeds, \\Leeds, LS2 9JT, UK\\
and\\
Landau Institute for Theoretical Physics, \\
Russian Academy of Sciences,\\
2 Kosygina st., Moscow, 117940, Russia\\ \\
V.V.~Sokolov\\ Centre for Nonlinear Studies, \\
at Landau Institute for Theoretical Physics,\\
Russian Academy of Sciences,
\\2 Kosygina st., Moscow, 117940, Russia}

\abstract{In this paper we give definitions of basic concepts
such as symmetries, first integrals, Hamiltonian and recursion operators
suitable for ordinary differential equations on associative algebras, and
in particular for matrix differential equations. We choose existence of 
hierarchies of
first integrals and/or symmetries as a criterion for integrability and
justify it by examples. Using our componentless approach we have solved
a number of classification problems for integrable equations on free 
associative algebras. Also, in the simplest case, we have listed
all possible Hamiltonian operators of low order.
}

\maketitle
\newpage

\section{Introduction}

In the classical theory (Lie, Liouville, etc.) of ordinary
differential equations (ODEs) there are remarkable results which
relate the property of integrability of ODEs in quadratures with the existence
of continuous symmetries and first integrals. Symmetries and first integrals
may serve as a solid mathematical foundation for an algebraic theory
of integrable equations. In the case of integrable partial differential 
equations (PDEs)
such a theory does already exist and proved to be extremely
efficient \cite{ss,msy,fokas,mss}. The key property of integrable PDEs
is the existence of infinite hierarchies of local infinitesimal
symmetries generated by a recursion operator. Characteristic
features of integrable Hamiltonian PDEs are multi-Hamiltonian
structures and hierarchies of local conservation laws. These
properties are well described in the fundamental monograph by
P.Olver \cite{olver} where references on original publications are
well presented.

A straightforward generalisation of these ideas to the case of
ODEs is impossible, since the number of independent commuting
symmetries and first integrals for a finite dimensional dynamical
system is finite and bounded by its dimension. To overcome this obstacle we
propose to study an intermediate object, namely, equations on free
associative algebras. Here we are going to show that such
equations are quite similar to PDEs. In particular they may have
infinite hierarchies of first integrals or symmetries and there
are plenty of reasons to choose these properties as an algebraic
definition of integrability.

If we are given a matrix differential equation with no
restrictions on the matrix dimension then we can treat it as an
equation on an abstract associative algebra. For example,
equations for the classical Euler top in $n$-dimensional space can
be written in a matrix form
\begin{equation}\label{eulerM}
\dot M=[M,\Omega] \, .
\end{equation}
Here $M,\Omega$ are real skew-symmetric $n\times n$ matrices of angular
momentum and angular velocity, brackets $[\, ,\, ]$ denote usual matrix
commutator and
\begin{equation}\label{omegam}
M=J\Omega+\Omega J\, ,
\end{equation}
where $J$ is a constant diagonal matrix of moments of inertia.
The problem of integration of the general
$n$ dimensional case was solved by Manakov \cite{man}. He made an important
observation that the Euler equation (\ref{eulerM}), (\ref{omegam})
is a stationary point of the $N$-wave equations, known to be integrable
\cite{nwave}.

Equation (\ref{eulerM}) is a Hamiltonian system\footnote{Indeed,
it can be rewritten in the form
$
\dot{M}=\{ M,H_1\} =[M,\mbox{grad}_M H_1]\,
$
with Poisson brackets defined as
$
\{F,G\}=\mbox{trace}(\mbox{grad}_M F,[M,\mbox{grad}_M G])
$
.} with Hamiltonian
\begin{equation}\label{ham}
H_1 =\frac{1}{2} \mbox{trace} (M\Omega )\, .
\end{equation}

It is easy to check that
\begin{equation}\label{H2}
H_2 =-\mbox{trace}(M^2 J^2)\, , \qquad
H_3 = \mbox{trace}(2J^4 M^2+ J^2 M J^2 M)
\end{equation}
are also first integrals of (\ref{eulerM}).
First integral $H_2$ generates a symmetry of the Euler equation
\begin{equation}\label{sym1}
M_\tau =[M,\mbox{grad}_M H_2]=J^2 M^2 -M^2 J^2\, .
\end{equation}

In terms of $M,J,\Omega$ first integrals and symmetries have
reasonably simple polynomial form. An attempt to find first
integrals or symmetries in component notations, say by the method
of indetermined coefficients, would fail for dimensions $n>4$
because of enormous scale of computations involved. In order to
avoid component-wise computations and make results suitable for
any dimensional problem we could regard objects $M,\Omega, J$ as
generators of an associative algebra (i.e. assume that all our
objects are polynomials of non--commutative variables $M,\Omega,
J$ with complex or real coefficients). This is not a free algebra
because of the constraint (\ref{omegam}). A more serious problem
is that the Euler equation (\ref{eulerM}) does not define
a $t$-derivation of the algebra. Indeed, we cannot determine
$\frac{d\Omega}{dt}$ as an element of the algebra using the Euler
equation  (\ref{eulerM}) and the constraint (\ref{omegam}). In
other words the Euler equation is not an {\sl evolutionary}
equation. On the contrary, symmetry (\ref{sym1}) with the condition
$J_\tau=0$ defines a derivation of a free algebra generated by
elements $M$ and $J$ (we shall denote 
associative multiplication in this algebra
by $\circ$). It is much easier to study
(\ref{sym1}) than the original Euler equation. Since the Euler
equation and (\ref{sym1}) belong to the same hierarchy, they have
common symmetries and first integrals.

In this paper we will use equation
\begin{equation}\label{mc}
M_{t}=M^2\circ C-C\circ M^2
\end{equation}
which coincides with (\ref{sym1}) if $C=J^2$, as a basic and
instructive example.

Being defined on a free algebra, equation (\ref{mc}) has infinite
hierarchies of polynomial symmetries and first integrals (see
Section 1) and therefore is integrable according to our
definition. Taking concrete finite and infinite dimensional
associative algebras (such as algebras of differential or integral
operators, matrix algebras or algebras with particular commutation
relations) instead of the universal free associative algebra one
can obtain a wide variety of equations which we expect to be
integrable in a more conventional sense. For instance, in matrix
realisations, symmetries and first integrals of equations on
abstract associative algebras are inherited by the corresponding
matrix ODEs and can be used for their integration in quadratures.

Our approach to differential equations on associative algebras
requires proper definitions for such concepts as symmetry, first
integral, Hamiltonian and recursion operators, etc. In the next
section we give a definition of symmetries and first integrals suitable for
equations on free associative algebras. Also we show that equation
({\ref{mc}) is not a unique representative of one component
integrable equations and formulate a few classification results
(we call them Results, rather than Proposition or Theorem, because
they were obtained by straightforward computations).

Fr\'echet  derivative, gradient, Hamiltonian and recursion
operators for equations on free associative algebras are defined
in Section 3. In the same section we formulate a simple
classification result for Hamiltonian operators. In particular, we
have shown that equation (\ref{mc}) is a tri--Hamiltonian
system. Ratios of the corresponding Hamiltonian operators yield
two recursion operators. Existence of two independent recursion
operators enables us to build up a complete two-index hierarchy of
symmetries for equation (\ref{mc}).

Many important {\sl multi--component} integrable equations
on associative algebras can be obtained as
reductions of {\sl one--component} equations. Here we give just one example.
Let $M$ and $C$ in (\ref{mc}) be represented by matrices of the form
\begin{equation}\label{MCmatr}
M=\left( \begin{array}{cccccc}
0&u_1&0&0&\cdot&0\\
0&0&u_2&0&\cdot&0\\
\cdot&\cdot&\cdot&\cdot&\cdot&\cdot\\
0&0&0&0&\cdot&u_{N-1}\\
u_{N}&0&0&0&\cdot&0
\end{array}\right)\, , \quad
C=\left( \begin{array}{cccccc}
0&0&0&\cdot&0&J_N\\
J_1&0&0&\cdot&0&0\\
0&J_2&0&\cdot&0&0\\
\cdot&\cdot&\cdot&\cdot&\cdot&\cdot\\
0&0&0&\cdot&J_{N-1}&0
\end{array}\right)\, ,
\end{equation}
where $u_k$ and $J_k$ are block matrices (of any dimension) or
even generators of a bigger free algebra. It follows from equation
(\ref{mc}) that $u_k$ satisfy the non-abelian Volterra equation
\begin{equation}\label{volter}
\frac{d}{dt}u_k=u_k \circ u_{k+1}\circ J_{k+1}-J_{k-1}\circ u_{k-1}\circ u_k,
\quad k\in\Integ _{N}\, .
\end{equation}
In Section 4 we consider other examples (modified Volterra
equation, discrete Burgers type equation, etc) and reductions to
two--component systems of equations. Also we solve a simplest
classification problem for two--component equations with one extra
symmetry and study some of their properties.

We conclude our paper by a discussion of possible developments and future
research.

\subsection{First integrals and symmetries for equations on free
associative algebras}

Equation (\ref{mc}) is naturally defined on a free algebra ${\cal
M}$ generated by elements $M$ and $C$ over the field $\Complex$
with an associative (but noncommutative) multiplication denoted as
$\circ$. Elements of the field will be denoted by Greek letters
$\alpha,\beta, \gamma \ldots \in \Complex$, elements of ${\cal M}$
will be denoted by Latin letters. Also we assume that ${\cal M}$
contains a unity element which we denote $ \, I\! d $.

All basic objects such as symmetries, first integrals, recursion
and Hamiltonian operators can be naturally defined for equations
on associative algebras. We formulate and illustrate definitions
for equations on algebra ${\cal M}$ generated by two elements $M$
(variable) and $C$ (constant), but all these definitions can be
extended to any associative algebra with a finite number of
generators straightforwardly.

\begin{Defi} For a differential equation $M_t=F$ on algebra ${\cal M}$
we say that $G\in{\cal M}$ is a generator of an infinitesimal symmetry
if the flows $M_{\tau}=G$ and $M_t=F$ commute
(i.e if $\frac{dG}{dt}$ evaluating according the equation and
$\frac{dF}{d\tau}$ evaluating
according $M_{\tau}=G$ give the same element of ${\cal M}$).
\end{Defi}

\noindent {\bf Remark.} In terms of the (non-commutative) differential
algebra, equation $M_t=F$ is nothing but a derivation $D_F$ of
algebra ${\cal M}$, which maps $M$ to $F$ and symmetry $M_{\tau}=G$
is another derivation $D_G$, commuting with $D_F$.
\vspace{.3cm}

To introduce the concept of first integrals, we need an analog of
trace, which is not yet defined in algebra ${\cal M}$. As the
matter of fact, in our calculations we use only two property of
the trace, namely linearity and a possibility to perform cyclic
permutations in monomials. Therefore, instead of a trace, we
define an equivalence relation
for elements of ${\cal M}$ in a standard way.

\begin{Defi} Two elements $f_1$ and $f_2$ of ${\cal M}$ will be called 
equivalent, denoted $f_1 \sim f_2$, iff $f_1$ can be obtained from $f_2$ by
cyclic permutations in its monomials.
\end{Defi}

For example, $\alpha M\circ C\circ M\sim \alpha C\circ M\circ M$, and
of course, any commutator is equivalent to zero.

\begin{Defi} Element $h$ of ${\cal M}$ will be called a first integral of
our differential equation, if $\frac{dh}{dt} \sim 0$.
First integrals $h_1$ and $h_2$ are said to be equivalent if $h_1 -h_2 \sim 0$.
\end{Defi}

Here there is an obvious similarity with a definition for conserved
densities in the theory of evolutionary PDEs \cite{olver}. In both cases,
first integrals and conserved densities are defined as elements of
equivalence classes, the
difference is in the choice of the equivalence relation.

According to our definition, $R_n=M^n\, , \ n=1,2, \ldots$ are
nontrivial first integrals of equation (\ref{mc}). Indeed,
$R_n\not\sim 0$ and $\frac{d}{dt}M^n=\sum_{k=1}^{n} M^{k-1}\circ
(M^2\circ C-C\circ M^2)\circ M^{n-k}$. It is easy to see that each
element of the sum is equivalent to zero.

The r.h.s. of equation  (\ref{mc}) is a (double) homogeneous
polynomial with respect to scaling $\sigma : C\rightarrow \mu C\,
,\ M\rightarrow \nu M$. For such type homogeneous polynomials the
exponents of $\mu$ and $\nu$ we shall call weights. As a linear
space, algebra ${\cal M}$ can be decomposed in a direct sum ${\cal
M}=\oplus {\cal M}_{nm}$ where $a\in {\cal M}_{nm}$ if $\sigma :
a\rightarrow \mu ^{n}\lambda ^{m} a$. It is easy to see that

\begin{Lem} Let $I$ be a first integral and $G$ a generator of a symmetry
for a (double) homogeneous equation. Then any (double) homogeneous component 
of $I$ or
$G$ (i.e. a projection on ${\cal M}_{nm}$) is also a first integral or
a generator of a symmetry respectively.
\end{Lem}

A direct computation of first integrals for (\ref{mc}) with small weights
is a simple problem, since we have only a few coefficients to determine.
For example a general homogeneous polynomial of weights $(2,2)$ is equivalent
to
\[
P_{2,2}=\alpha C^2\circ M^2+\beta C\circ M\circ C\circ M\, .
\]
It is easy to check that $\frac{dP_{2,2}}{dt}\sim 0$ iff $\alpha=2\beta$
and therefore
\begin{equation}\label{I22}
I_{2,2}=2C^2\circ M^2+ C\circ M\circ C\circ M
\end{equation}
is a first integral of equation (\ref{mc}).

In a similar way one can easily find that apart from the obvious orbit
integrals $I_{0,n}=M^n$, the following polynomials $I_{n,1}=C^n\circ  M\, ,
I_{1,n}=C\circ  M^n$ and
\begin{equation}\label{I32}
I_{3,2}=C^3\circ M^2+C^2\circ M\circ C\circ M\, \qquad
I_{2,3}=C^2\circ M^3+C\circ M\circ C\circ M^2,
\end{equation}
are nontrivial first integrals of equation (\ref{mc}). It is interesting
to note that the space of first integrals in each homogeneous component
${\cal M}_{kn}$ is one dimensional. We have found the above integrals
(\ref{I22}) and (\ref{I32}) by a simple straightforward computation.
We did not use the known Lax representation for equation (\ref{mc}) (
not known for all equations), but if we do so, we would find that
$I_{n,m}$ is equivalent to the projection of $(M+C)^{(n+m)}$ on
${\cal M}_{nm}$.

A similar computation gives us generators of homogeneous symmetries 
$S_{n,m}\in {\cal M}_{nm}$ for (\ref{mc}).
For instance, to obtain $S_{2,2}$ we  start with a general
ansatz
\[
Q_{2,2}=\alpha _1 C^2\circ M^2+\alpha _2 C\circ M\circ C\circ M+
\alpha_3 C\circ M^2\circ C+\alpha_4 M\circ C^2\circ M+
\alpha_5 M\circ C\circ M\circ C+\alpha_6 M^2\circ C^2.
\]
It is easy to verify that $Q_{2,2}$ gives a symmetry for  (\ref{mc}) iff
$\alpha_3=\alpha_4=0$ and  $\alpha_1=\alpha_2=-\alpha_5=-\alpha_6$.

Having first integrals and symmetries of equation (\ref{mc}) we
can use them for the integration of Euler's equation. Let $C=J^2$,
with $J$ being a diagonal matrix with positive eigenvalues, and let $M$ be
a skew-symmetric real matrix. Then one can show that
$H_{n,m}=\mbox{trace} (I_{n,m})$ is a first integral of the Euler
equation. Of course, only half of the integrals survive such a
reduction, since $M$ is a skew symmetric matrix and therefore
$H_{n,2k-1}=0$. Also, the first integrals may become functionally
dependent. For instance, in the case of the three dimensional
Euler top only two first integrals are independent. We can choose
$H_{0,2}$ and $H_{1,2}$ as a basic set of functionally independent
first integrals. In the four dimensional case as a basic set we
can choose $H_{0,2},H_{1,2},H_{2,2}$ and $H_{0,4}$. In both cases
the Hamiltonian (\ref{ham}) of the Euler equation does not belong
to the basic set (it is a function of the basic integrals).

Summarising the above exercise we would like to say the following: The
Euler top on an abstract associative algebra is not an
evolutionary equation and therefore it is difficult, if possible
at all, to study its symmetries and first integrals directly. Meanwhile, it
has an evolutionary symmetry (\ref{mc}) which is naturally defined
on a free algebra. It is very easy to find first integrals and
symmetries for (\ref{mc}) (only linear algebraic equations 
to be solved). In a standard matrix case, symmetries and first
integrals of (\ref{mc}) are also symmetries and first integrals of
the Euler equation, they are sufficient for integration of the
Euler equation in quadratures.

\subsection{Other examples of integrable evolutionary equations on ${\cal M}$.}

Equation (\ref{mc}) possesses sequences of first integrals and
symmetries. These properties enable us to integrate its matrix
(finite dimensional) reductions. A natural question is: whether there
exist other equations which possess such properties? In this
section we show that the answer is affirmative. The definitions
given above proved to be efficient and suitable for handling such problems.

Let us inquire when a general double homogeneous evolutionary equation on
${\cal M}$:
\begin{equation}\label{gen3}
M_t =\alpha M^2\circ C+\beta  M\circ C\circ M+\gamma C\circ  M^2\,
\end{equation}
possesses symmetries? According to Lemma 1, we can study symmetries
in each homogeneous component independently. General equation
(\ref{gen3}) has an obvious sequence of symmetries $S_{1,n}=M
\circ C^n-C^n\circ M\, ,n=1,2,\ldots$. In the matrix case these
symmetries would correspond to the invariance of our equation with
respect to similarity transformations commuting with $C$.
Symmetries $S_{1,n}$ we shall call trivial and omit them from the
further consideration.

We checked the existence of homogeneous symmetries $S_{n,m}\in {\cal
M}_{n,m}$ with $n+m\le 10$ for equation (\ref{gen3}) by
straightforward computations. To do so, we wrote a general
homogeneous polynomial $P_{n,m}\in {\cal M}_{n,m}$ with
indetermined coefficients, and commuted equation (\ref{gen3})
with flow $M_{\tau}=P_{n,m}$. Then we equated the result of the
commutation to zero and solved the resulting system of algebraic
equations for the coefficients of $P_{n,m}$ and $\{
\alpha,\beta,\gamma \}$. This system is nonlinear (all equations
are quadratic), but quite simple. All steps in this study are
algorithmic. In order to perform actual computations we wrote a
symbolic code for Mathematica.

\begin{Res} Up to scaling of $t$, every equation of the form (\ref{gen3})
possessing a nontrivial symmetry $S_{n,m}\in{\cal M}\, ,n+m\le 10$
coincides with one of the following equations:
\begin{eqnarray}
M_t&=&[M^2 , \,C]\, ,\label{man}
\\ M_t&=&[M, \, M\circ C]\,
,\label{ska}\\ M_t&=&[ C\circ M, \,M]\, ,\label{skb}\\
M_t&=&M\circ C\circ M\, .\label{mcm}
\end{eqnarray}
\end{Res}

The first equation of the list coincides with (\ref{mc}).
Equations (\ref{ska}) and (\ref{skb}) are equivalent in the
following sense. Let us define a formal involution $\star$ on
${\cal M}$ by
\begin{equation}\label{star}
 M^\star=M, \ \ C^\star =C,\ \ (A\circ B)^\star =
B^\star \circ A^\star,\ \ (\alpha A+\beta B)^\star=\alpha A^\star
+\beta B^\star , \ \ A,B\in {\cal M}, \ \ \alpha,\beta\in
\Complex.
\end{equation}
If an equation $M_t=P$ has a symmetry $M_{\tau}=S$ then equation
$M_t=P^\star $ has symmetry $M_{\tau}=S^\star$. The r.h.s. of
equation (\ref{skb}) is a result of the involution $\star $
applied to (\ref{ska}). Similarly, all results related to
(\ref{ska}) can be easily rearranged for (\ref{skb}).

The form of symmetries for equations (\ref{ska}), (\ref{mcm}) that we
found with the help of our computer programme has given us
a hint for the following statement, which can be easily proven.

\begin{Prop}\label{propsym} Equation (\ref{ska}) has commuting symmetries
\begin{equation}\label{symska1}
M_{\tau _{n,m}}=[M,\, M^{n-1}\circ C^m]\, ,\ \ \ n,m\in \Nat\, .
\end{equation}
Equation (\ref{mcm})  has commuting symmetries
\begin{equation}\label{symmcm}
M_{\tau _{2,m}}=M\circ C^{m}\circ M\, ,\ \ \ m\in \Nat\, .
\end{equation}
\end{Prop}

Another interesting problem is to describe equations of the form
(\ref{gen3}) which possess first integrals. Straightforward
computations lead to the following

\begin{Res}
Equation (\ref{gen3}) has first integrals of the form
$I_{n,0}=M^n$ iff $\alpha +\beta +\gamma=0$. If equation
(\ref{gen3}) has other first integrals $I_{n,m}, \,
n+m\le 10$, then up to scaling of $t$ it coincides with
(\ref{man}).
\end{Res}

Equation (\ref{mcm}) does not have first integrals, and equation
(\ref{ska}) has only obvious orbit integrals, but both equations
have a hierarchy of symmetries. It indicates the
``C-integrability'' of (\ref{mcm}) and (\ref{ska}) (a similar
situation for PDEs is described in \cite{ss}).

Indeed, if $M,C$ are $n\times n$ matrices, then a solution of the
Cauchy problem ($M(0)=M_0$) for equation (\ref{mcm}) is
\cite{jord}:
\[ M(t)=({\bf e} -M_0 C t)^{-1} M_0 \, ,\]
where ${\bf e}$ is the unit matrix.

Equation (\ref{ska}) in the matrix case can be integrated in the
following way. Let $M_0$ be any constant matrix and $T$ be a
solution of the following linear differential equation with
constant coefficients
\begin{equation}\label{Tt}
 T_t =M_0 TC \, ,\qquad T|_{t=0}={\bf e}
\end{equation}
then $M(t)=T^{-1}M_0 T$ satisfies equation (\ref{ska}) and $M(0)=M_0$.
Moreover, if $T(t,\tau_{n,m})$ is
a simultaneous fundamental solution for equation  (\ref{Tt}) and
$T_{\tau_{n,m}}=M_0^n T C^m$ (these equations are obviously compatible), then
$M(t,\tau_{n,m})=T^{-1}M_0 T$ satisfies
equation $M_{\tau_{n,m}}=[M,M^{n-1}C^m]$ as well.

Equations (\ref{man}) and (\ref{ska}) have cubic (in $M$) symmetries.
The next natural question is: whether there exist other
cubic double homogeneous equations
\begin{equation}\label{gen4}
M_t =\alpha M^3\circ C+\beta M^2\circ C\circ M+\gamma M\circ C\circ  M^2+
\delta C\circ M^3\,
\end{equation}
which possess symmetries or first integrals. The answer can be formulated as
follows.

\begin{Res}
Up to scaling of $t$ and formal involution $\star$, every equation of
the form (\ref{gen4})
possessing a nontrivial symmetry $S_{n,m}\in{\cal M}\, ,n+m\le 10$
coincides with one of the following equations:
\begin{eqnarray}
M_t&=&[M^3 ,\, C]\, ,\label{symman}\\ M_t&=&[M,\, C\circ M^2]\,
,\label{symska}\\ M_t&=&[M\circ C\circ M,\, M]\, ,\label{mines1}\\
M_t&=&M^3\circ C-M^2\circ C\circ M+M\circ C\circ M^2-C\circ
M^3,.\label{plus2}
\end{eqnarray}
\end{Res}

The first two equations (\ref{symman}) and (\ref{symska}) we expected, 
since they are
symmetries of equations (\ref{man}) and (\ref{ska}) respectively.

Equation (\ref{plus2}) is new, and has symmetries with weights $(2k-1,1)$,
but does not possess symmetries of weights $(2k,1)$. It is interesting to
note that element $N=M^2$ satisfies the equation $N_t =[N^2,C]$ (\ref{man}).
In the scalar case we would consider these equations to be point-equivalent.
In the matrix case if we know solutions of (\ref{plus2}) we know solutions of
equation (\ref{man}) as well, but not vice-versa.

Moreover, we can find an equation for the ``$n$--th root'' of $N$. Indeed,
it is easy to show that if $M$ satisfies equation
\begin{equation}
M_t=M^{n+1}\circ C-M^{n}\circ C\circ M+M\circ C\circ M^{n}-C\circ
M^{n+1},.\label{plusn}
\end{equation}
then $N=M^{n}$ satisfies equation (\ref{man}). Equation (\ref{plusn})
has symmetries with weights $(kn+1,1)\, , k=0,1,\ldots $, but does not have
symmetries of weights $(p,1)$ where $p\not \equiv 1(\mbox{mod}(n))$.

In the above list, equation (\ref{mines1}) is maybe the most
interesting one. It has symmetries of weights $(n,1)$ for any $n$.
This equation also can be related on a formal level with an
extension of the hierarchy of equation (\ref{man}). If algebra
${\cal M}$ contains (or is extended by) the inverse element
$N=M^{-1}$, then it follows from (\ref{mines1}) that
$N_{t}=[N^{-1},C]$ and it is a symmetry of equation $N_\tau
=[N^2,C]$ (\ref{man}). Generally speaking, the inverse element may
not exist (for example it does not exist in the $3\times 3$ matrix
case when  $M$ is skew--symmetric).

In a similar way one can verify that equation
\begin{equation} \label{minesn}
M_t=[M,\, M^{n}\circ C\circ M^{n}] \label{minusn}
\end{equation}
is related with (\ref{mines1}) by the substitution $N=M^{n}$.
Similar to (\ref{plusn}), this equation has lacunas of length $n$
in the sequence of symmetries, namely it has symmetries of weights
$(kn+1,1)\, , k=2,3,\ldots$.

\section{Hamiltonian and recursion operators on ${\cal M}$.}

\subsection{Basic definitions}
In this section we give definitions for some basic objects such as
the Fr\'echet and variational derivatives, Hamiltonian operators,
etc. suitable for equations on free algebra ${\cal M}$. We shall
use equation (\ref{mc}) to illustrate our definitions.

For any $a\in{\cal M}$ we define operators of left ($L_a$) and right
($R_a$) multiplications which map ${\cal M}$ into itself according
the following rule: let $b$ be any element of ${\cal M}$, then
\[
  L_a (b)=a\circ b, \ \ \  R_a (b)=b\circ a\, .
\]
It follows from the above definition and associativity of algebra
${\cal M}$ that $R_a L_b=L_b R_a$ and
\begin{equation}\label{LR}
L_{a\circ b} =L_a L_b, \ \ R_{a\circ b} =R_b R_a, \ \
L_{\alpha a+\beta b}=\alpha L_a+\beta L_b,\
\ R_{\alpha a+\beta b}=\alpha R_a+\beta R_b.
\end{equation}

Let us define an algebra  ${\cal O}$  with generators $L_M,L_C,R_M,R_C$
satisfying the following relations:
\[
R_M L_M=L_M R_M,\ \ R_M L_C=L_C R_M,\ \ R_C L_M=L_M R_C,\ \ R_C L_C=L_C R_C\, .
\]
We shall call ${\cal O}$ as the algebra of local operators. Due to (\ref{LR})
any operator of multiplication on elements of ${\cal M}$ can be represented
by a corresponding element of ${\cal O}$. Let us denote by $id$ the unity element
of ${\cal O}$, i.e. the operator of multiplication by $ \, I\! d $.

The gradation of ${\cal M}$ induces a gradation structure
$ {\cal O}=\oplus {\cal O}_{n,m}$ on ${\cal O}$. If we scale $C\rightarrow \mu C$ and
$M\rightarrow \nu M$, then elements of ${\cal O}_{n,m}$ gain the multiplier
$\mu ^n \nu^m$. And, of course, if $A\in {\cal O}_{n,m}$, then
$A:{\cal M}_{p,q}\rightarrow {\cal M}_{p+n,q+m}$. we shall call 
elements of ${\cal O}_{n,m}$
homogeneous operators of the weight $(n,m)$.

The Fr\'echet derivative $a_*$ for any element  $a\in{\cal M}$ belongs to
${\cal O}$.

\begin{Defi} Let $a=a(M,C)$ be any element of ${\cal M}$.  Then the Fr\'echet
derivative $a_{*}\in {\cal O}$ of $a$ is uniquely defined by:
\[ \frac{d}{d\epsilon}a(M+\epsilon\, \delta\! M,C)|_{\epsilon =0}=
a_{*}(\delta \! M)\, .\]
\end{Defi}

For example, following this definition we can calculate the Fr\'echet
derivative $F_{*}$ of element $F(M,C)=M^2\circ C-C\circ M^2$ which is the
r.h.s. of equation (\ref{mc}):
\begin{eqnarray*}
&& \frac{d}{d\epsilon}(F(M+\epsilon \, \delta \! M),C))|_{\epsilon =0}=
\frac{d}{d\epsilon}((M+\epsilon \, \delta \! M)^2\circ C-
C\circ (M+\epsilon \, \delta \! M)^2)|_{\epsilon =0}=\\
&&\delta \! M\circ M \circ C+M\circ \delta \! M\circ C-
C\circ \delta \! M\circ M -C\circ M\circ \delta \! M=F_{*}(\delta \! M)
\end{eqnarray*}
Therefore
\begin{equation}\label{frechF}
F_{*}=R_C R_M+L_M R_C-L_C R_M - L_C L_M
\end{equation}

Let $D_{F}$ and $D_{G}$ be two derivations of algebra ${\cal M}$
corresponding to elements $F,G\in {\cal M}$ respectively (cf. Remark
after the Definition 1). Then their commutator is also a
derivation of the algebra ${\cal M}$
\[ D_{K}=[D_{F},\, D_{G}]\, , \]
with
\[  K=G_{*}(F)-F_{*}(G) \, .\]
In particular, $G$ is a generator of a symmetry for equation
$M_t=F$ iff
\begin{equation}\label{symeq}
G_{*}(F)=F_{*}(G)\, .
\end{equation}
Usually equation (\ref{symeq}) is called the defining equation for
symmetries (c.f. \cite{olver}).

Any derivation $D_{F}$ of ${\cal M}$ induces a derivation
of ${\cal O}$:
\begin{equation}\label{LRt}
D_{F}L_{a}=L_{a_{\star}(F)},\ \ \ \ \ D_{F}R_{a}=R_{a_{\star}(F)}
\end{equation}
for any $a\in{\cal M}$.

The formal involution $\star $ (\ref{star}) induces an involution of ${\cal O}$:
\begin{equation}\label{starL}
L^\star _a =R_{a}\,
\end{equation}
It follows from (\ref{starL}) that  $R^\star _a =L_{a}$ and
in particular $R^\star _M =L_M,\ \ L^\star _C =R_C$.
Operator $Q\in {\cal O}$ is called symmetric or skew--symmetric
if $Q^\star =Q$ or $Q^\star =-Q$ respectively. For example
operator $ad_M=L_M-R_M$ is skew-symmetric. It is useful to remember
that for any $a,b\in {\cal M}$ and $Q\in {\cal M}$
\[  a\circ Q(b)\sim  Q^{\star}(a) \circ b .\]

\subsection{Hamiltonian structures on ${\cal M}$.}

The general Hamiltonian equation on ${\cal M}$ has the form
\begin{equation}\label{hform}
M_{t}=\Theta (\mbox{grad}_M (H(M,C)))
\end{equation}
where $H(M,C)$ is a Hamiltonian of the equation, $\Theta$ is a
Hamiltonian operator. Here we shall study local Hamiltonian
operators, i.e. assume that $\Theta \in {\cal O}$.

For any  $a\in{\cal M}$ we define the gradient $\mbox{grad}_M(a)\in {\cal M}$
as follows:

\begin{Defi} Let $a(M,C), \delta \! M\in{\cal M}$. Then
$ \mbox{grad}_M(a(M,C))$ is uniquely defined by:
\[ \frac{d}{d\epsilon}a(M+\epsilon \, \delta \! M,C)|_{\epsilon =0}\sim
\delta \! M\circ \mbox{grad}_M(a(M,C))\, .\]
\end{Defi}

It is easy to check that if $H_1 \sim H_2\, ,H_i\in{\cal M}$ then
$ \mbox{grad}_M(H_1)=\mbox{grad}_M(H_2)$ and in particular if
$H_1\sim 0$ then $\mbox{grad}_M(H_1)=0$. Moreover, the following
analog\footnote{The above defined $\mbox{grad}_M$ is an analog of
the variational derivative in the theory of PDEs.} of the well known Theorem
(cf. \cite{gmsh}) holds.

\begin{Prop} Let $a\in {\cal M}$, then $a\sim \mbox{const}\, ,
\mbox{const}\in {\cal M}$ iff $\mbox{grad}_M(a)=0$.
\end{Prop}

For example, let us verify that equation (\ref{mc}) can be written
in a Hamiltonian form (\ref{hform}) with Hamiltonian
$H(M,C)=C\circ M^2$ and Hamiltonian operator $\Theta=ad_M$ ($\Theta
(a)=M\circ a-a\circ M$ for any $a\in {\cal M}$). It is enough to
show that $\mbox{grad}_M(H(M,C))=M \circ C +  C \circ M$. Indeed,
\[ H(M+\epsilon \, \delta \! M,C)=H(M,C)+
\epsilon C \circ \delta \! M \circ M +\epsilon C \circ M \circ \delta \! M+O(\epsilon^2) \]
and
\[ \frac{d}{d\epsilon}H(M+\epsilon \, \delta \! M,C)|_{\epsilon=0}=
C \circ \delta \! M \circ M + C \circ M \circ \delta \! M\sim
\delta \! M \circ (M \circ C +  C \circ M)=\delta \! M \circ\mbox{grad}_M(H(M,C)) \,
.\]

\begin{Defi} We shall call $\Theta\in {\cal O}$ a Hamiltonian operator, if
the Poisson bracket
\[  \{a,b\}=\mbox{grad}_M a\circ  \Theta ( \mbox{grad}_M b)  ,\ \ \ \
a,b\in {\cal M} \]
satisfies conditions
\begin{equation}\label{skewsymm}
\{a,b\}+\{b,a\}\sim 0
\end{equation}
\begin{equation}\label{jacoby}
\{a,\{b,c\}\}+\{b,\{c,a\}\}+\{c,\{a,b\}\}\sim 0
\end{equation}
for any elements $a,b,c\in {\cal M}$.
\end{Defi}

With the help of the {\sl substitution principle} \cite{olver} one can show that
(\ref{skewsymm}) implies
\begin{equation}\label{skewtheta}
\Theta ^{\star}=-\Theta
\end{equation}
(i.e. $\Theta $ is a skew--symmetric operator with respect to the involution
(\ref{starL})).   Moreover, the Jacoby identity (\ref{jacoby}) is equivalent
to the following condition (same as (7.11) on page 428 in \cite{olver}):
\begin{equation}\label{thjac}
Q_2 \circ X_{3}(Q_1)+Q_3\circ X_{1}(Q_2)+Q_1\circ
X_{2}(Q_3)\sim 0,
\end{equation}
where operators $X_{i}\in {\cal O}$ are defined as follows
\[X_{i}=D_{\Theta (Q_{i})}(\Theta)\, .\]

It is easy to check that $\hat{\Theta}=ad_{C^k}=L_{C^k}-R_{C^k}$ is a
Hamiltonian
operator for any $k$. Indeed, it is a skew--symmetric operator
and satisfies (\ref{thjac}) since $X_{i}=0$. Operator
$\Theta_1=ad_{M}=L_{M}-R_{M}$ also satisfies conditions (\ref{skewtheta}),
(\ref{thjac}), and therefore, is a Hamiltonian operator on ${\cal M}$.
Hamiltonian operators $\hat{\Theta}$ and $\Theta_1$ are homogeneous and have
weights $(k,0)$ and $(0,1)$ respectively.
For any $\lambda\in \Real$, operator
\begin{equation}\label{thlam}
\Theta(\lambda)=\Theta_1+\lambda\hat{\Theta}
\end{equation}
is a Hamiltonian operator as well. In other words, operators
$\hat{\Theta}$ and $\Theta_1$
are compatible and form a Hamiltonian pencil $\Theta(\lambda)$. As usual,
the Hamiltonian pencil can be obtained by a simple shift: if we replace
$M$ by $M+\lambda C^k$ in $\Theta_1$, we find
\[
ad_{M+\lambda C^k}=L_{M+\lambda C^k}-R_{M+\lambda C^k}=
\Theta_1+\lambda\hat{\Theta} \, .
\]

We can employ the method of indetermined coefficients for
searching Hamiltonian operators of relatively low weights.
Conditions (\ref{skewtheta}) and (\ref{thjac}) yield systems of
linear and quadratic equations for the coefficients of $\Theta$.

\begin{Res} Up to scaling of $M$ and $C$, every homogeneous Hamiltonian
operator of weights $(0,m),$ $ m<8$ and $(1,m),$  $m<7$ coincides
with one of the following:
\begin{eqnarray}
\Theta_0&=&R_C-L_C\, ,\label{T0}\\
\Theta_1&=&R_M-L_M\, ,\label{T1}\\
\Theta_2&=&L_C R_M-L_M R_C\, ,\label{T2}\\
\Theta_3&=&L_M^2 R_M-L_M R_M^2\, ,\label{T3}\\
\Theta_4&=&L_M L_C L_M R_M-L_M R_M R_C R_M\, .\label{T4}
\end{eqnarray}
\end{Res}

The shifts $C \rightarrow C+\lambda  \, I\! d $ and $M \rightarrow M+\mu  \, I\! d $
in  $\Theta_2$ gives a Hamiltonian
pencil $\Theta_2+\lambda \Theta_1+\mu \Theta_0$ for the Euler top and its
hierarchy (i.e. equation (\ref{mc}), for example). Thus equation (\ref{mc})
is a {\sl three Hamiltonian} system.

A similar shift $C \rightarrow C+\lambda  \, I\! d $ of $\Theta_4$
gives a pencil $\Theta_4+\lambda \Theta_3$. If we assume the
element $M$ to be invertible, then the following formal change of variables
$\tilde M=M^{-1}$ relates operators $\Theta_1$ and $\Theta_2$ with
$\Theta_3$ and $\Theta_4$ respectively.

\subsection{Recursion operators.}

Recursion operators, if they exist, give a convenient way for
generating a hierarchy of symmetries (see, for instance,
\cite{olver}). For PDEs, most of known recursion operators are
pseudo-differential. Existence of local, i.e. differential,
recursion operators usually indicates on C--integrability of the
equation \cite{mss}. Here we begin with a definition of local
recursion operators for ODEs on ${\cal M}$ and later extend it
to a ``non local'' case.

\begin{Defi}

We say that $\Lambda \in {\cal O}$ is a recursion operator for
an evolutionary equation
\begin{equation}\label{evol}
M_t=F,\qquad M\in {\cal M},
\end{equation}
if it satisfies the following equation:
\begin{equation}\label{recur}
\Lambda_t=F_*  \Lambda-\Lambda F_*\, ,
\end{equation}
where $F_{*}$ is the Fr\'echet derivative of $F$

\end{Defi}

Operators satisfying equation (\ref{recur}) act on the
space of symmetries of equation (\ref{evol}).

\begin{Prop}
If $\Lambda$ is a recursion operator and $G$ is the generator of
a symmetry $M_{\tau }=G$ for equation (\ref{evol})
then $G_{1}=\Lambda (G)$
is also a generator of a symmetry $M_{\tau_{1} }=G_{1}$ for equation (\ref{evol}).
\end{Prop}

\noindent
{\bf Proof.}
Indeed, let us check that $(M_t)_{\tau_{1}}$ and $(M_{\tau_{1}})_t$ 
represent the same element of ${\cal M}$. We have: 
$(M_t)_{\tau_{1}} =
F_* \Lambda (G)$ and
\[ (M_{\tau_{1}})_t =\Lambda _t(G)+\Lambda (G_t)=
(F_*  \Lambda-\Lambda F_* )(G)+\Lambda G_* F= \Lambda (G_*
F-F_{*}G)+F_*  \Lambda \, .\] Now, due to the condition
(\ref{symeq}), we have $(M_t)_{\tau_{1}} =(M_{\tau_{1}})_t.$
$\Box$

If we have two operators $\Lambda_1$ and $\Lambda _2$ satisfying equation
(\ref{recur}), then any linear combination
$\alpha\Lambda_1+\beta\Lambda _2$ with constant coefficients $\alpha,\beta\in
\Complex$, a composition $\Lambda_1 \Lambda _2$ and, in particular,
any power $\Lambda_1 ^{n}$  also satisfy (\ref{recur}).

Applying
the sequence $\Lambda^{k},\ k=1,2,\ldots$ to a single symmetry (or to
equation (\ref{evol}) itself) one could generate a hierarchy of symmetries.
In the case of integrable one--component evolution PDEs
the whole hierarchy of symmetries is generated by a single recursion
operator. In our case of equations on ${\cal M}$ the situation
is rather different. As we have seen above (cf. (\ref{symska1})), integrable
equations may have two-index
hierarchy of symmetries. It gives us a hint that we should have
{\sl two} independent recursion operators, and each
of the operators raises the corresponding index.

For example, $\Lambda=L_M$ is a local recursion operator for
equation (\ref{ska}). Indeed, according (\ref{ska}), $\frac{d}{dt}\Lambda=
L_{M^2 C}-L_{MCM}$. The Fr\'echet derivative of the r.h.s. in this case is
\[ F_*= R_{MC}+L_M R_C-R_{CM}-L_{MC} \]
and therefore
\[ F_*  \Lambda-\Lambda F_*=(R_{MC}+L_M R_C-R_{CM}-L_{MC})L_M-
L_M (R_{MC}+L_M R_C-R_{CM}-L_{MC})=L_{M^2 C}-L_{MCM} \, .\]
The following sequence
$M_{\tau_{1,n+2}}=L_{M^n} (M^2 C-MCM)=M^{n+2}C-M^{n+1}CM,\, n\in \Nat$ is
an infinite
hierarchy of symmetries for equation (\ref{ska}). It is not the whole set
of symmetries (Proposition \ref{propsym}, (\ref{symska})) and we shall see
that a ``non local'' recursion operator is responsible for the
remaining part of the hierarchy.

In the theory of integrable PDEs a non local recursion operator, by
definition, is a ratio of two local (i.e differential) operators.
In our case we define the corresponding object as a ratio
of two operators from ${\cal O}$.

One of possible ways to introduce non local recursion operators
for multi--Hamiltonian equations belongs to Magri \cite{magri}:

\begin{Prop}\label{MAG}
If $\Theta$ and $\Theta_{1}$ are two Hamiltonian operators for the same
equation, then their ratio
\begin{equation}\label{ththm1}
\Lambda=\Theta_1 \Theta^{-1}
\end{equation}
is a recursion operator.
\end{Prop}

\noindent
{\bf Proof.}
Indeed, any Hamiltonian operator satisfies equation
(c.f. (7.38), page 449, \cite{olver})
\begin{equation}\label{hamil}
\Theta_t=F_* \Theta+\Theta F_* ^{\star}\, .
\end{equation}
If $\Theta$ and $\Theta_{1}$ are two solutions of equation
(\ref{hamil}), then it is easy to check that the ratio
$\Lambda=\Theta_1 \Theta^{-1}$ satisfies equation (\ref{recur}).
$\Box$

In order to illustrate this Proposition let us consider equation (\ref{mc}).
It has three Hamiltonian operators $\Theta_0= L_C-R_C=ad_C $
(\ref{T0}), $\Theta_1=L_M-R_M =ad_M$ (\ref{T1}) and
$\Theta_2=L_M R_C-L_C R_M$ (\ref{T2}). Consequently, we have two independent
recursion operators for (\ref{mc}):
\begin{equation}\label{l12mc}
\Lambda_{1}=(L_M R_C-L_C R_M) ad_{M}^{-1}\, ,\qquad
\Lambda_{2}=(L_M R_C-L_C R_M) ad_{C}^{-1}\, .
\end{equation}

Here we propose a simple generalisation of Proposition \ref{MAG},
and make it suitable even for non--Hamiltonian equations.

\begin{Prop}\label{MY}
Let  $Q_{1},Q_{2}\in {\cal O}$ be two solutions of
operator equation
\begin{equation}\label{eqQ}
Q_t=F_* Q+Q P\, ,
\end{equation}
where $P\in {\cal O}$ is a given operator. Then
\begin{equation}\label{lamq21}
\Lambda=Q_{2}Q_{1}^{-1}
\end{equation}
 is a recursion operator for equation
(\ref{evol}).
\end{Prop}

\noindent
{\bf Proof.} Indeed,
\[
\Lambda_{t}=\frac{dQ_{2}}{dt}Q_{1}^{-1}-Q_{2}Q_{1}^{-1}\frac{dQ_{1}}{dt}Q_{1}^{-1}=
(F_* Q_{2}+Q_{2} P)Q_{1}^{-1}-Q_{2}Q_{1}^{-1}(F_* Q_{1}+Q_{1}
P)Q_{1}^{-1}=F_{*}\Lambda-\Lambda F_{*}\, .
\]
$\Box$

Equation (\ref{eqQ}) is a generalisation of two well known operator
equations. Namely, if $P=-F_*$, then (\ref{eqQ}) coincides with equation
(\ref{recur}) for a recursion operator. If $P=F_* ^{\star},$ then
skew-symmetric solutions of (\ref{eqQ}) give us Hamiltonian operators
(\ref{hamil}). Here we would like to emphasise that in Lemma \ref{MY}
we do not specify the nature of the operator $P$, the only requirement
is an existence of two (or more) solutions $Q_1$ and $Q_2$. Such a construction 
may occur to be useful for theory of integrable PDEs as well.

It is clear that for a homogeneous equation (\ref{evol}) with
$F\in {\cal M}_{n,m}$, operator $P$ must belong to ${\cal
O}_{n,m-1}$ and we could try to find simultaneously operators
$P,Q_{1}$ and $Q_{2}$ by the method of indetermined coefficients.
Actually, following this simple idea we have found recursion
operators for equations (\ref{ska}),(\ref{mcm}) and
(\ref{mines1}).

Equation (\ref{mcm}) is not a Hamiltonian system, but it has a hierarchy
of symmetries. It is easy to find that
\[ Q_n=L_M L_{C}^{n}\, ,\ \ \ \ n=0,1,\ldots \]
are solutions of equation (\ref{eqQ}) with $P=R_M R_C$. Therefore
\begin{equation}\label{lmcm}
\Lambda = L_M L_C (L_M)^{-1}
\end{equation}
 is a non local recursion operator
for (\ref{mcm}).

In a similar way one can show that besides of a local recursion
operator $\Lambda_{1}=L_M$, equation (\ref{ska}) has a non local
recursion operator $\Lambda_2 =ad_M R_C ad_M^{-1}$ which
corresponds to $Q_{1}=ad_{M}, Q_{2}=ad_{M} R_{C}$ and
$P=L_{MC}-L_{M}R_{C}$

It is easy to check that $\Theta_3$ (\ref{T3}), $\Theta_4$ (\ref{T4})
are Hamiltonian operators for equation $M_t=[M\circ C\circ M,M]$
(\ref{mines1}). Therefore
\begin{equation}\label{lam43}
 \Lambda_{1}= \Theta_4 \Theta_{3}^{-1}= (L_M L_C L_M R_M-L_M R_M R_C R_M)
(L_M^2 R_M-L_M R_M^2)^{-1}=(L_M L_C- R_M R_C)ad_M^{-1}
\end{equation}
is a recursion operator for (\ref{mines1}). In the last part of the equality
(\ref{lam43}), i.e. after a cancellation, we have a ratio of $Q_1=L_M L_C- R_M
R_C$ and $Q_2=ad_M$. Neither of these two factors is a Hamiltonian
operator for (\ref{mines1}), but they both satisfy the same
equation (\ref{eqQ}),
where  $P=L_M L_C R_M-L_M R_M R_C$ is a local operator.
Moreover, $Q_3= L_M R_M R_C-L_M L_C R_M$ is a third independent 
solution of (\ref{eqQ}).
Therefore $\Lambda_{2}=Q_3 ad_{M}^{-1}$ is another
recursion operator for equation (\ref{mines1}).

In spite of a non local recursion operator is formally defined by
expression (\ref{lamq21}) in Proposition \ref{MY}, its action
on symmetries is not well defined yet. Here there are two
problems:
\begin{itemize}
\item Action of non local operator
(\ref{lamq21}) can be correctly defined only on elements of 
the image space ${\cal M}_{Q_{1}}\subset {\cal M}$  of operator $Q_{1}$.
\item  If $Q_{1}$ has a non--trivial kernel,
then action of $\Lambda$ is not uniquely defined.
\end{itemize}
Similar picture we had in the theory of integrable PDEs \cite{olver}, where
often there was a problem to define the action of
$D_{x}^{-1}$ on differential polynomials.

As a first example we consider an action of recursion operator
(\ref{lmcm}) on symmetries of equation (\ref{mcm}). 
In this case the kernel space
of operator $Q_{1}=L_{M}$ is trivial. The action of $Q_{1}^{-1}$ 
(and therefore of $\Lambda$)
is correctly defined on subset ${\cal M}_{Q_{1}}=\{M\circ a;a\in {\cal
M}\}$ by $Q_{1}^{-1}(M\circ a)=a$. Powers $\Lambda^{k}=L_{M}C^{k}L_{M}^{-1}$
of the recursion operator $\Lambda$ are also correctly defined on ${\cal
M}_{Q_{1}}$ and the r.h.s. of equation
(\ref{mcm}) belongs to ${\cal M}_{Q_{1}}$. Therefore we can
generate the following infinite hierarchy of symmetries
\[ S_{n,2}=\Lambda ^{n-1} (M\circ C\circ M)=M\circ C^n \circ M \, ,\quad
n=0,1,2,\ldots \ \]
for equation (\ref{mcm}). Equation (\ref{mcm}) has other
symmetries, namely the trivial symmetry $M_{\tau_{0}}=M\circ
C-C\circ M$ and the scaling symmetry $M_{s_{1}}=M+tM\circ
C\circ M$. Action of $\Lambda$ (\ref{lmcm}) on the generator of
the trivial symmetry is not correctly defined (i.e. the result
does not belong to ${\cal M}$). On the contrary, the action of
$\Lambda^{k}$ on the generator of the scaling symmetry is
correctly defined and give the following hierarchy of time
dependent symmetries
\[ M_{s_{k}}=\Lambda^{k-1}(M+tM\circ
C\circ M)=M\circ C^{k-1}+tM\circ C^{k} \circ M \, ,\quad
k=0,1,2,\ldots \, .\]
As usual (c.f. \cite{fokas}), time independent part of 
the above hierarchy defines a
hierarchy of master symmetries. Namely,
\[
S_{m+n,2}=S_{m,2*}(K_{n})-K_{n*}(S_{m,2})\, ,
\]
where
$K_{n}=M\circ C^{n}$ are generators of master symmetries.

In the case when operator $Q_{1}$ (see Proposition \ref{MY}) has a
nontrivial kernel we have to impose some extra conditions in order
to make the action of $Q_{1}^{-1}$ uniquely defined. As an
example, let us consider equation (\ref{mc}) with recursion
operator $\Lambda_{1}$ (\ref{l12mc}).
We can choose $M^k\, , k=1,2,\ldots$ as a basis in the kernel
space of $ad_M$.  Applying the recursion operator $\Lambda_{1}$
to the trivial symmetry generated by $S_{1,1}=M \circ C-C\circ M$ we obtain
\begin{eqnarray*}
G&=&\Lambda_{1} (S_{1,1})=\Theta_2 (C)+
\sum \alpha_k \Theta_2 (M^k)\\
&=&M\circ C^2-C^2\circ M^2+
\sum \alpha_k (M^{k+1}\circ C-C\circ M^{k+1})\, .
\end{eqnarray*}
We see that accounting elements of the kernel space we add
symmetries $S_{1,k}=M^{k}\circ C-C\circ M^{k}\in {\cal M}_{1,k}$
to $G$. The function $S_{1,1}$, operators $\Theta_1=ad_{M},
\Theta_2=L_{M}R_{C}-L_{C}R_{M}$ are
homogeneous and if we request that the function $G=S_{2,1}=\Lambda_{1}
(S_{1,1})$ to be homogeneous too, we have to choose coefficients
$\alpha_k=0$. It is easy to see that $S_{2,1}$ is an element from the
image space of $ad_M$,
and therefore $S_{3,1}=\Lambda_{1} (S_{2,1})=\Lambda_{1}^2(S_{1,1})$
is well defined and is the next
member in the  hierarchy of symmetries of equation (\ref{mc}), etc.
Moreover, it is easy to check, that elements $S_{k,1}$ belong to
the image of another Hamiltonian operator $\Theta_{0}=ad_{C}$ and
therefore the action of operator $\Lambda_{2}$ (\ref{l12mc}) on
these generators is also correctly defined. The whole two--index
hierarchy of symmetries for equation (\ref{mc}) can be obtained by
the action of recursion operators $\Lambda_{n,m}=\Lambda_{2}^{n}\Lambda_{1}^{m}$ on
the trivial symmetry
\begin{equation}\label{lnm}
S_{n,m}=\Lambda_{m-1,n-1}(S_{1,1})\, .
\end{equation}
The same hierarchy of symmetries can be
obtained using the first integrals $I_{n,m}$
\[ S_{n,m}=ad_M \mbox{grad}_M (I_{n,m})=\Theta_2 \mbox{grad}_M (I_{n-1,m})\, .\]

In a similar way one can check that pairs of recursion operators
for equations (\ref{ska}) and (\ref{mines1}) give corresponding
two--index hierarchies of symmetries.

\section{Multicomponent equations}
\subsection{Breeding and reducing equations}

In the previous sections we were dealing with evolutionary equations on
a free algebra ${\cal M}$ with one constant ($C$) and one variable ($M$)
generators. Our result indicates (see Result 1) that there are only three
basic hierarchies of equations and one could think that it may not be worth
to develop a sophisticated theory to serve these three exceptional cases.
In this section we are going to demonstrate that many important integrable
equations are nothing but particular cases of these three equations.

A way to breed equations is to regard $M$ and $C$ as $N\times N$
or even infinite dimensional matrices whose entries are generators
of a free algebra ${\cal A}$ and to reduce the system obtained by
imposing linear constrains on the entries  compatible with the dynamics.

One example which gives the (non-abelian) Volterra equation
(\ref{volter}) has been considered in Intro\-duc\-tion. Deriving
equation (\ref{volter}) from (\ref{mc}) we have made two steps: we
have replaced $M,C$ by $N\times N$ matrices whose entries belong
to a free algebra and then we have imposed constrains by setting
some of the entries equal to zero. These constrains are compatible
with the dynamics. Indeed, let $C$ be given and be of the form
(\ref{MCmatr}). If the initial conditions are such that matrix $M$
has the form (\ref{MCmatr}) then it will remain of the same form
for any $t$.  The constrains imposed are not compatible with all
the symmetries $S_{n,m}$ of equation (\ref{mc}). Only symmetries
$S_{n,n+1}$ survive under the reduction. First integrals of
equation (\ref{volter}) can be obtained from the first integrals $I_{n,m}$
of (\ref{mc}) by taking a formal trace of the matrix corresponding
to  $I_{n,m}$
(i.e. summing up diagonal elements). For example
\begin{eqnarray*}
\rho_1&=&\mbox{Tr}(I_{1,1})=\mbox{Tr}(M\circ C)=
\sum_{n\in\Integ _{N}}J_n \circ u_n\, ,\\
\rho_2&=&\mbox{Tr}(I_{2,2})=\mbox{Tr}(2M^2\circ C^2+M\circ C\circ M\circ
C)=\\&=&
\sum_{n\in\Integ _{N}}J_n \circ u_n\circ J_n \circ u_n+
2 J_n \circ u_n \circ u_{n+1}\circ J_{n+1}\, .
\end{eqnarray*}
It is easy to see that $\mbox{Tr}(I_{n,m})=0$ if
$n\not\equiv m\, \mbox{mod}(N)$.

The modified Volterra equation
\begin{equation}\label{mvolter}
\frac{d}{dt}u_k=u_k \circ (J_{k-1}\circ u_{k-1}-u_{k+1}\circ J_{k})\circ u_k,
\quad k\in\Integ _{N}\, .
\end{equation}
is a reduction of equation (\ref{mines1}), where we have to assume $M$ to
be of the form (\ref{MCmatr}) and matrix $C$ is of the form
$[C]_{p,q}=\delta_{p+2,q}^{N} J_q$ (here $\delta_{i,j}^N =1$ if $i\equiv j\,
\mbox{mod}(N)$ and $0$ otherwise). Again, symmetries and first integrals
of equation (\ref{mvolter})  can be easily obtained from the
corresponding symmetries and first integrals of equation (\ref{mines1}).

Cyclic reduction (\ref{MCmatr}) is also compatible with equation (\ref{ska}).
The corresponding C--integrable system of equations for variables $u_k$ is:
\begin{equation}\label{Cvolter}
\frac{d}{dt}u_k=u_k \circ u_{k+1}\circ J_{k+1}-u_{k}\circ J_{k}\circ  u_k,
\quad k\in\Integ _{N}\, .
\end{equation}

Let us consider the simplest nontrivial case $N=2$ and assume that
$J_1=J_2= \, I\! d $.
In this case equations (\ref{Cvolter}) and (\ref{volter}) are reduced to
\begin{equation}\label{Cvol2}
u_t=u\circ u-u\circ v\, ,\quad v_t=v\circ v-v\circ u\, ,
\end{equation}
and
\begin{equation}\label{vol2}
u_t=u\circ v-v\circ u\, ,\quad v_t=v\circ u-u\circ v\, ,
\end{equation}
respectively. One more two component equation can be obtained from
(\ref{volter}) if we assume $N=3,\, J_1=J_2=J_3= \, I\! d $ and
$u_3=-u_1 -u_2$:
\begin{equation}\label{vol3}
u_t=u\circ u+u\circ v+v\circ u\, ,\quad v_t=-v\circ v-u\circ v-v\circ u\, .
\end{equation}

All these evolutionary equations are defined on a free algebra ${\cal A}$
with generators $u,v$ over the field $\Complex$. Elements $u$ and
$v$ satisfy a system of equations of the form
\begin{equation}\label{uv}
u_t=P(u,v)\, ,\quad v_t=Q(u,v)\, ,
\end{equation}
where $P(u,v)$ and $Q(u,v)$ are elements of  ${\cal A}$. The involution
$\star$ (\ref{star}) on ${\cal A}$ is defined by
\begin{equation}\label{star1}
u^\star=u\, ,\quad v^\star=v\, ,\quad (a\circ b)^\star=
b^\star \circ a^\star \, ,\quad a,b\in {\cal A}.
\end{equation}

Equations, which are related to each other by linear
transformations of the form
\begin{equation}\label{invert}
\hat{u}=\alpha u+\beta v\, ,\quad \hat{v}=\gamma u+\delta v\, ,\quad
\alpha \delta-\beta\gamma \ne 0
\end{equation}
and involution (\ref{star1}), we shall call {\sl equivalent}.

Equations which are equivalent to
\begin{equation}\label{treug}
u_t=P(u,v)\, ,\quad v_t=Q(v)\,
\end{equation}
we shall call {\sl triangular}.

For example, (\ref{vol2}) is a triangular equation.
Indeed, after the following change of variables
$\hat{u}=u,\hat{v}=u+v$ we obtain $ \hat{u}_t= [\hat{u},\hat{v}],
\, \hat{v}_t=0 \, .$ It is easy to check that equation
(\ref{vol3}) is equivalent to
\begin{equation}\label{v2u2}
u_t=v^2\, ,\quad v_t=u^2\,
\end{equation}
and not triangular. Equations (\ref{vol3}) and (\ref{Cvol2}) are not
equivalent.

\subsection{The simplest classification problem for equations on algebra
${\cal A}$}
\subsubsection{Quadratic equations with a cubic symmetry}

Equations (\ref{Cvol2}), (\ref{vol3}) have infinite hierarchies of symmetries
and first integrals. It is interesting to answer the question whether do
exist other quadratic equations
\begin{equation}
\begin{array}{l}
u_t   =  \alpha_1 u \circ u  + \alpha_2 u \circ v + \alpha_3  v \circ u +
\alpha_4 v \circ v,\\
v_t   =  \beta_1 u  \circ u  + \beta_2 u  \circ v + \beta_3  v \circ u + \beta_4
v \circ v \label{equ}
\end{array}
\end{equation}
which possess symmetries or first integrals? And, if so, how many classes
of inequivalent and non--triangular equations do exist? A partial answer to
these questions is given by

\begin{Theor}
Any non-triangular equation (\ref{equ}) possessing
 a symmetry of the form
\begin{eqnarray} \nonumber
u_{\tau}   &=&\gamma _1 u\circ u\circ u  + \gamma _2 u\circ u\circ v +
\gamma _3 u\circ v\circ u + \gamma _4 v\circ u\circ u  +  \\ \nonumber
&&\gamma _5 u\circ v\circ v  + \gamma _6 v\circ u\circ v +
\gamma _7 v\circ v\circ u + \gamma _8 v\circ v\circ v ,\\ \nonumber
v_{\tau} & = &\delta _1 u\circ u\circ u  + \delta _2 u\circ u\circ v + \delta _3
u\circ v\circ u + \delta _4 v\circ u\circ u  + \\ \nonumber
&&\delta _5 u\circ v\circ v  + \delta _6 v\circ u\circ v +
\delta _7 v\circ v\circ u + \delta _8 v\circ v\circ v ,
\end{eqnarray}
is equivalent to one of the following:

\begin{eqnarray}
&&\begin{array}{l}
u_t=u\circ u - u\circ v,\\
v_t=v\circ v - u\circ v + v\circ u,     \label{eq1}
\end{array}
\\&&
\begin{array}{l}
u_t=u\circ v,\\
v_t=v\circ u,       \label{eq2}
\end{array}
\\&&
\begin{array}{l}
u_t=u\circ u - u\circ v,\\
v_t=v\circ v - u\circ v,        \label{eq3}
\end{array}
\\&&
\begin{array}{l}
u_t= - u\circ v,\\
v_t=v\circ v + u\circ v - v\circ u,     \label{eq4}
\end{array}
\\&&
\begin{array}{l}
u_t=u\circ v - v\circ u,\\
v_t=u\circ u + u\circ v - v\circ u,     \label{eq5}
\end{array}
\\&&
\begin{array}{l}
u_t=v\circ v,\\
v_t=u\circ u,       \label{eq6}
\end{array}
\end{eqnarray}
\end{Theor}

It is a remarkable fact, that a requirement of existence of just
one cubic symmetry selects a finite list of equations with no free
parameters (or more precisely, all possible parameters can be
removed by linear transformations (\ref{invert})).

The next natural question is whether equations
(\ref{eq1})-(\ref{eq6}) have other symmetries, do they possess
first integrals, whether the corresponding matrix equations are
integrable? We have calculated all
time independent polynomial homogeneous symmetries and first
integrals of low orders for equations of the list.
The dimensions of linear spaces of symmetries and
first integrals are compiled in the following
table: \vspace{.5cm}

\noindent
\begin{tabular}{|c|cccccccc|cccccccccccc|}\hline
&\multicolumn{8}{c|}{Number of Symmetries}&
\multicolumn{12}{|c|}{Number of First Integrals} \\ \hline
Order               &1&2&3&4&5&6&7&8&1&2&3&4&5&6&7&8&9&10&11&12\\ \hline
Equation (\ref{eq1})&0&1&1&0&0&0&0&0&0&0&0&0&0&0&0&0&0&0&0&0\\
Equation (\ref{eq2})&0&1&1&1&1&1&1&1&1&1&1&1&1&1&1&1&1&1&1&1\\
Equation (\ref{eq3})&0&1&1&1&1&1&1&1&0&1&0&1&0&1&0&1&0&1&0&1\\
Equation (\ref{eq4})&0&1&1&1&0&1&0&1&0&1&0&1&0&2&0&2&1&3&2&6\\
Equation (\ref{eq5})&0&1&1&1&0&1&0&1&1&1&1&2&2&4&4&8&11&20&27&52\\
Equation (\ref{eq6})&0&1&1&0&2&1&1&2&0&0&1&1&0&2&1&1&2&2&1&3\\
\hline
\end{tabular}
\vspace{.5cm}

None of the equations from the list has a linear symmetry. The only symmetry of
order two is the equation itself.

Equation (\ref{eq1}) has a symmetry of order 3
\begin{equation}
\begin{array}{l}
u_{\tau}   =  u \circ  u \circ v\, ,\\
v_{\tau}   = v \circ u \circ v-u \circ v \circ v\, .
\end{array}
\end{equation}
(and that is the reason why it belongs to the list) but it does not have any
higher
symmetries nor any first integrals (at least of orders less than nine and
thirteen
respectively).

Equation (\ref{eq2}) has a hierarchy of symmetries
\begin{equation}
\left( \begin{array}{l}
u_{\tau_n}\\
v_{\tau_n}  \end{array}\right)=
 \Lambda ^n    \left( \begin{array}{l}
u\circ v\\ v\circ u \end{array}\right)=\left( \begin{array}{l}
u \circ  v \circ ( u-v)^n\\
v \circ u \circ (u-v)^n
\end{array}\right)\, ,\quad n=0,1,2,\ldots
\end{equation}
where the corresponding recursion operator $\Lambda$ is of the form
\[
\Lambda=\left( \begin{array} {ccc} R_u - R_v, & 0\\
0, & R_u - R_v
\end{array} \right)\, .
\]
It is easy to verify that
\[ \rho _n=(u-v)^n \]
is a sequence of first integrals for equation (\ref{eq2}).
Equation (\ref{eq2}) can be written in a Hamiltonian form
\[
\left( \begin{array} {ccc} u_t\\
v_t
\end{array} \right) =\Theta \left( \begin{array} {ccc} \mbox{grad}_u \rho_1\\
\mbox{grad}_v \rho_1
\end{array} \right)
\]
where the Hamiltonian operator $\Theta$ is of the form
\[
\Theta=\left( \begin{array} {ccc} R_u R_u-L_u L_u, & L_u L_v+L_u R_v-
L_v R_u+R_u R_v\\
L_u R_v-L_v L_u - L_v R_u-R_v R_u, & L_v L_v - R_v R_v
\end{array} \right)\, .
\]
It is interesting to note that equation (\ref{eq2}) is a stationary point of
1+1 dimensional integrable system \cite{golsok}
\begin{eqnarray*}
u_t&=& u_x + u \circ v\, ,\\
v_t&=& - v_x + v \circ u\, .
\end{eqnarray*}

Equations (\ref{eq3}) and (\ref{eq2}) have different sequences of orders of
first integrals -
it is an independent indication that these two equations are not equivalent,
i.e.
cannot be related by invertible transformation (\ref{invert}) and involution
(\ref{star1}).
It is easy to check that
\[ \rho_{2n}=(u\circ v)^n \]
is a sequence of first integrals for equation (\ref{eq3}). 
Its symmetries also can be written in
terms of a recursion operator
\begin{equation}
\left( \begin{array}{l}
u_{\tau_{2n}}\\
v_{\tau_{2n}}  \end{array}\right)=
 \Lambda ^n    \left( \begin{array}{l}
-u\circ v\\ v\circ v+u\circ v -v\circ u \end{array}\right)\, ,\quad
n=0,1,2,\ldots
\end{equation}

\begin{equation}
\left( \begin{array}{l}
u_{\tau_{2n+1}}\\
v_{\tau_{2n+1}}  \end{array}\right)=
 \Lambda ^n    \left( \begin{array}{l}
u \circ u \circ v  - u \circ v \circ u  \\
 -u \circ v \circ v  + v \circ u\circ v
 \end{array}\right)\, ,\quad n=0,1,2,\ldots
\end{equation}
where
\[
 \Lambda =\left( \begin{array} {ccc} R_u R_v - L_u R_v, & L_u R_u - L_u L_u \\
R_v R_v - L_v R_v,  & L_u R_v - L_v L_u
\end{array} \right)=
\]
\[
=\left( \begin{array} {ccc} R_u - L_u, & 0\\
0, & R_v-L_v
\end{array} \right)
\left( \begin{array} {ccc} R_v, & L_u\\
R_v, & L_u
\end{array} \right)\, .
\]

Equation (\ref{eq6}) is, maybe, the most interesting in the list. It is a
Hamiltonian equation
with
\[
\Theta =\left( \begin{array} {ccc} 0, & 1\\
-1, & 0
\end{array} \right)
\]
and Hamiltonian $H={1\over 3} (v^3-u^3).$ We have seen in the
previous section that (\ref{eq6}) is equivalent to (\ref{vol3}) which is
a reduction of equation (\ref{mc}). Therefore the Lax representation for
(\ref{eq6}) is known
and its symmetries and first integrals can be easily found. A straightforward
attempt to
find a local recursion operator (i.e. to find solution of equation
(\ref{recur})) gives
\begin{equation}\label{lam3}
 \Lambda=\left( \begin{array} {ccc} L_u L_v - L_u R_v-L_v R_u+R_u R_v,
& -L_u L_u +2 L_u R_u - R_u R_u \\
L_v L_v - 2 L_v R_v+R_v R_v,  & L_u R_v - L_v L_u+L_v R_u-R_v R_u
\end{array} \right)=
\]
\[
=\left( \begin{array} {ccc} L_u - R_u, & 0\\
0, & L_v-R_v
\end{array} \right)
\left( \begin{array} {ccc}L_v- R_v, & -L_u+R_u\\
L_v-R_v, & -L_u+R_u
\end{array} \right)\, .
\end{equation}
Equation itself and all known to us time independent symmetries belong to the
kernel of this ``recursion'' operator $\Lambda$ (\ref{lam3}). Equation (\ref{eq6}) is
homogeneous and therefore it has a scaling symmetry
\[ u_{\tau_s}=u+t v^2\, ,\quad v_{\tau_s}=v+t u^2 \, .\]
Applying $\Lambda$ to this scaling symmetry we obtain a cubic symmetry
\begin{equation}
\left(\begin{array}{l}
u_{\tau} \\v_{\tau}
\end{array}\right)  = \Lambda \left(\begin{array}{l}
u+t v^2 \\v+t u^2
\end{array}\right) =\left(\begin{array}{r}
- 2 u \circ u \circ v + 4 u \circ v \circ u -2 v \circ u \circ u\\
    2u \circ v \circ v  - 4 v \circ u \circ v +2 v \circ v \circ u
\end{array}\right)
\end{equation}
of equation (\ref{eq6}).

A systematic description of symmetries and first integrals for
equations (\ref{eq4}) and (\ref{eq5}) is still an open problem.

\subsubsection{Quadratic equations with a quartic symmetry}

Integrable equation (\ref{Cvol2}) does not possess a cubic symmetry and
therefore does not belong to the list (\ref{eq1})-(\ref{eq6}), but it has
a quartic symmetry
\begin{eqnarray}
&&\begin{array}{l}u_\tau=
u  \circ v\circ u^2 -u  \circ v^2  \circ  u\, ,\\
v_\tau=v\circ u  \circ v ^2 - v\circ u^2  \circ v\, .
\end{array}\label{eq444}
\end{eqnarray}
Symmetry (\ref{eq444}) and higher order symmetries of equation
(\ref{Cvol2}) can be generated by the following recursion
operator:
\[
 \Lambda=\left( \begin{array} {ccc} L_u L_v,
0 \\
0,  & L_v L_u
\end{array} \right) \, .
\]

\begin{Prop}\label{p24} Apart from (\ref{eq2})-(\ref{eq5}) the
list of quadratic equations on ${\cal A}$ which possess a quartic symmetry
includes the following
inequivalent equations:
\begin{eqnarray}
&&\begin{array}{l}u_t= u  \circ  u-v  \circ  u\, ,\\
v_t=v  \circ v- u  \circ v\, ;\end{array}\label{eq44}\\
&&\begin{array}{l}u_t=- u  \circ v\,
 ,\\v_t=v  \circ v+ u  \circ v\, ;\end{array}\label{eq41}\\
&&\begin{array}{l}u_t=-v  \circ  u\,
,\\v_t=v  \circ v+ u  \circ v\, ;\end{array}\label{eq42}\\
&&\begin{array}{l}u_t= u  \circ  u- u  \circ v-2 v  \circ  u\, ,\\
v_t=v  \circ v-2  u  \circ v-v  \circ  u\, ;\end{array}\label{eq45}\\
&&\begin{array}{l}u_t= u  \circ  u-2 v  \circ  u\, ,\\
v_t=v  \circ v-2 v  \circ  u\, ;\end{array}\label{eq48}\\
&&\begin{array}{l}u_t= u  \circ  u-2  u  \circ v\, ,\\
v_t=v  \circ v+4 v  \circ  u\, .\end{array}\label{eq49}
\end{eqnarray}
\end{Prop}

Here we do not claim that the list presented above is complete. Some
extra work is required to cast Proposition \ref{p24} into the form
similar to Theorem 1.

It is interesting to look at the sequence of dimensions for symmetries and  first integrals
of equations (\ref{eq44})-(\ref{eq49}).
\vspace{.5cm}

\noindent
\begin{tabular}{|c|cccccccc|cccccccccccc|}\hline
&\multicolumn{8}{c|}{Number of Symmetries}&
\multicolumn{12}{|c|}{Number of First Integrals} \\ \hline
Order                 &1&2&3&4&5&6&7&8&1&2&3&4&5&6&7&8&9&10&11&12\\ \hline
Equation (\ref{eq44}) &0&1&0&1&0&1&0&1&0&1&0&1&0&1&0&1&0&1&0&1\\
Equation (\ref{eq41}) &0&1&0&1&0&1&0&1&0&1&0&1&0&1&0&1&0&1&0&1\\
Equation (\ref{eq42}) &0&1&0&1&0&1&0&1&0&1&0&1&0&1&0&1&0&1&0&1\\
Equation (\ref{eq45}) &0&1&0&1&0&1&1&1&0&0&0&1&0&1&0&1&1&1&0&2\\
Equation (\ref{eq48}) &0&1&0&1&1&1&1&1&0&0&1&1&0&2&1&1&2&2&1&3\\
Equation (\ref{eq49}) &0&1&0&1&0&0&0&0&0&0&0&0&0&0&0&0&0&0&0&0\\
\hline
\end{tabular}
\vspace{.5cm}

Equations (\ref{eq44}),  (\ref{eq41}) and  (\ref{eq42}) have
identical sequences for dimensions of symmetries and first
integrals. Nevertheless we have checked that they are not
equivalent, i.e. cannot be related by linear transformations
(\ref{invert}) and involution (\ref{star1}).

Equation (\ref{eq49}) possesses a symmetry of order four but it seems it
does not possess any other symmetries or first integrals.

\section*{Instead of a conclusion}

In this paper, we have made an attempt to bring the rich hidden structure
of integrable PDEs \cite{mss} to a new domain, namely, to differential
equations on free associative algebras (see also \cite{olsok1,olsok2,balsok}).
We have formulated basic definitions
and shown their correctness and efficiency. Like integrable PDEs,
ODEs on associative algebras may have infinite hierarchies of
symmetries and first integrals and that asserts an algebraic
definition of integrability. In the case of finite
dimensional (matrix) representations of the algebra, the
corresponding (matrix) systems of ODEs inherit these symmetries and first
integrals and can be integrated in quadratures.

This study raises a lot of questions and open entirely new area for research.
For instance,
we foresee that the list of integrable equations on associative
algebras with constrains should be a lot bigger. Quantum problems,
with some commutation relations, naturally fall in this class.
Another important and promising
problem for further research is a systematic study
of reductions of equations on associative algebras and, in particular
corresponding matrix systems of ODEs. We expect that infinite dimensional
realisations (for instance by operators in a Hilbert space) and their
reductions may be of interest as well. We would like, also, to look
at the theory of classical integrable tops
from the point of such a componentless approach. That would require
to study equations on associative algebras with a few constant elements
related by algebraic constrains.

\section*{Acknowledgements} 

The first author (A.M.) was supported, in part, by  the Royal Society.
The second author (V.S.) was supported, in part, by RFBR grant 99-01-00294,
the INTAS project and EPSRC grant GR/L99036.
He is grateful to Leeds University (UK) for its hospitality.

\end{document}